\documentclass{elsart}
\usepackage[dvips]{graphics}
\usepackage{amsmath}
\usepackage{amsbsy}
\usepackage{amssymb}
\usepackage{epsfig}
\newcommand{\be}{\begin{equation}}
\newcommand{\ee}{\end{equation}}

\begin{document}

\begin{frontmatter}

\title{Comparison between methods for the determination of the primary
cosmic ray mass composition from the longitudinal profile of 
atmospheric cascades}

\author[1]{M.Ambrosio}, \author[1]{C.Aramo\corauthref{cor1}}
\ead{aramo@na.infn.it}, 
\author[1,2,3]{C.Donalek}, \author[4]{D.D'Urso}, 
\author[1,4,5]{A.D.Erlykin},  \author[1,2]{F. Guarino}, 
\author[4]{A.Insolia}, \author[1,2,6] {G.Longo}

\address[1]{INFN, Napoli Unit, Napoli, Italy}
\address[2]{ Department of Physical Sciences, University of Napoli
Federico II, Napoli, Italy}
\address[3]{Department of Math. and Appl., University
of Napoli Federico II, Napoli, Italy}
\address[4]{Department of Physics and Astronomy, University of Catania and 
INFN, Catania, Italy}
\address[5]{ P. N. Lebedev Physical Institute, Moscow, Russia}
\address[6]{INAF, Napoli Unit, Napoli, Italy}
\corauth[cor1]{Corresponding author}

\begin{abstract}

The determination of the primary cosmic ray mass
composition from the longitudinal development of atmospheric
cascades is still a debated issue. In this work we discuss several
data analysis methods and show that if the entire information
contained in the longitudinal profile is exploited, reliable
results may be obtained. Among the proposed methods 
FCC ('Fit of the Cascade Curve'), MTA ('Multiparametric Topological
Analysis') and NNA ('Neural Net Analysis') with conjugate gradient 
optimization algorithm give the best accuracy.
\end{abstract}
\end{frontmatter}
\section{Introduction}

The study of the longitudinal profile of individual
atmospheric cascades started in the early eighties with the
development of the fluorescent light detection technique
implemented for the first time within the framework of the Fly's
Eye experiment \cite{Baltrusaitis}. After these
pioneering efforts, there has been only one experimental array
continuing this type of studies \cite{Matthews} and
only recently a new and much more powerful detector has started to
collect data: the Fluorescent Detector (FD) of the Pierre Auger
Observatory \cite{Auger}. This instrument will produce a
large data flow over the next decades and is therefore calling for
new and accurate data analysis procedures capable to fully exploit 
the large amount of information contained in the FD data.\\
It is rather surprising, that while there are many methods, both parametric
and non-parametric (KNN, Bayesian methods, pattern recognition,
neural nets etc.), used to discriminate individual cascades on the
basis of ground--based information \cite{Haungs}, 
very little has been done to
exploit the amount of information contained in FD data. To
our knowledge, in fact, the most popular method developed so far makes use
of the depth of the maximum cascade development ($X_{max}$)
\cite{Gaisser} and derives the observed {\em mean} mass
composition as a function of the primary energy. 
Since in experiments which use fluorescent light for the study of the
longitudinal development of atmospheric cascades and for the determination
of their energy there is a minimum bias in 
the detection of cascades of different origin, the
observed mass composition coincides practically with the primary
composition. It has also to be stressed that this approach relies
on statistical grounds and therefore does not allow the
identification of the primary particle for each individual
cascade. Furthermore, even though in the longitudinal profile the
$X_{max}$ parameter is the most sensitive to the mass of the
primary particle, its sensitivity is still weak. For instance, at
a primary energy of 1 EeV (10$^{18}$eV) the mean iron induced
cascade has $X_{max}$ only (11-12)\% lower than for a proton
induced one, i.e. a difference which is of the same order of
magnitude as the intrinsic fluctuations in $X_{max}$. \\
As we shall discuss below, such unsatisfactory situation
improves drastically if other, seemingly less significant
parameters, are taken into account. Among them, we might have:
$N_{max}$ - the number of particles (mostly electrons) in the
maximum of the cascade, the speed of rise in the particle number
etc. For instance, at fixed primary energy, $N_{max}$ is about the
same for all cascades, and even though iron induced cascades
produce more muons and less energy is carried out by the
electrons, the effects on $N_{max}$ are very small. However, due
to the lower energy per constituent nucleon in the primary iron nucleus,
the cascade development and the rise of the cascade curve are on
average faster than for cascades originating from protons. 
This useful information is neglected when only $N_{max}$ is taken into
account. This type of arguments triggered our efforts to find
methods which make use of a larger amount of information
contained in the cascade curves and which are capable
to allow the identification (at least in terms
of probabilities) of the cascade origin also for individual
showers. Another statistical approach which uses the whole information on the 
longitudinal profile of the shower for the identification of its origin was 
developed by Risse M. et al. \cite{Risse}.\\
In what follows we shall focus on data similar to those
expected from the Fluorescent Detector of the Pierre Auger
Observatory: namely on the longitudinal profile of each atmospheric
cascade, i.e. on the number of charged particle $N_{ch}$ as the function of 
the atmospheric depth $X$. As it will be shown below, this
profile carries more information than $X_{max}$ or $N_{max}$ alone.\\
We wish to stress that even though the fraction of the hybrid
events, in which the information on the shower from the Pierre Auger
Surface Array is supplemented by the Fluorescent Detector data,
will hardly exceed 10\% of the total statistics accumulated by the Surface 
Array, these events need to be properly handled since they contain the maximum
information. In this paper we restrict ourselves to the
analysis of the longitudinal development of cascades.
Although tailored for possible applications in the context of the Pierre 
Auger experiment, the methods described below are quite general and may find 
application in other similar experiments.
\section{The simulated data}
\begin{figure}[t]
\begin{center}
\includegraphics[height=14cm,angle=0]{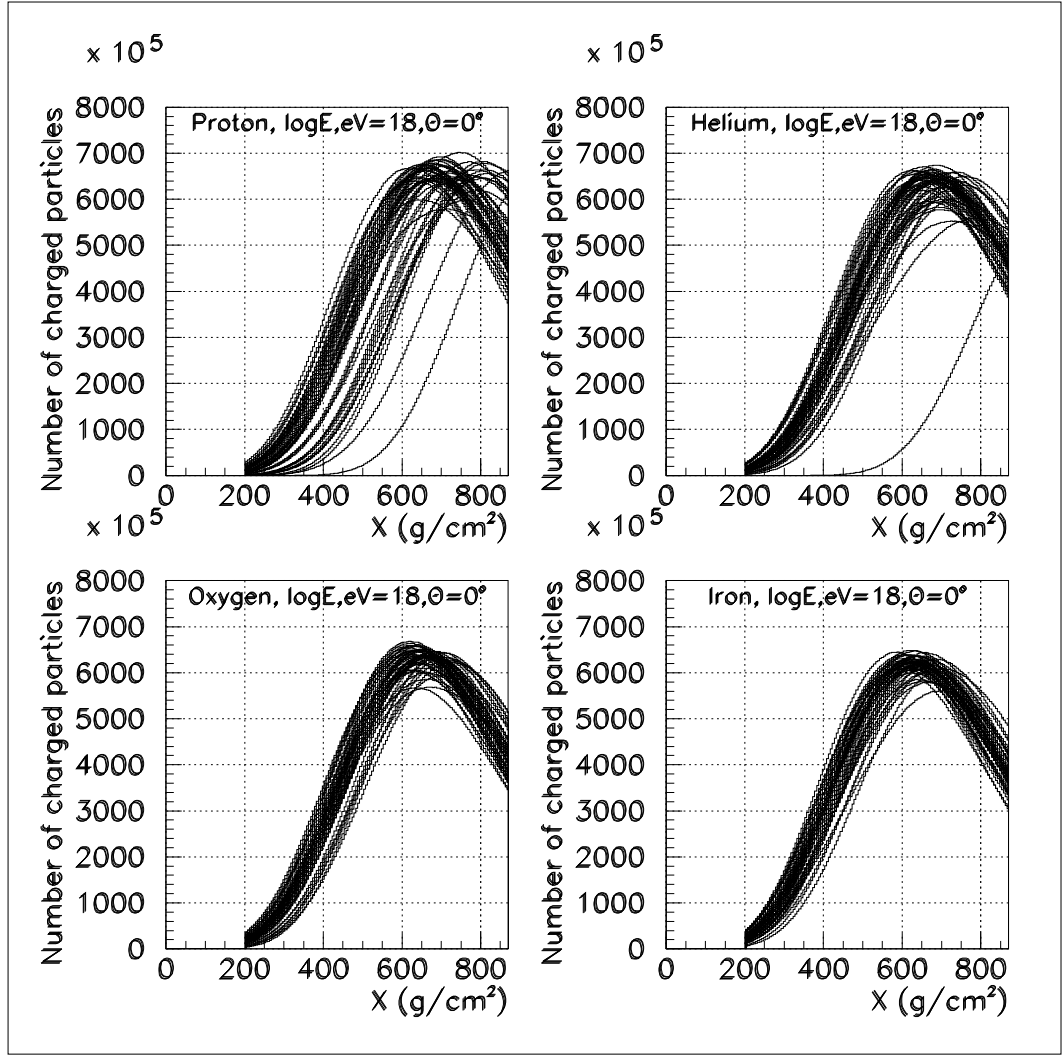}
\caption{\footnotesize A few examples of the longitudinal
development of 1 EeV vertical atmospheric cascades induced by
(clockwise from the upper left panel) protons P, helium He, oxygen
O and iron Fe nuclei, respectively. Each plot shows 50 typical
cascades.\label{fig:fcc0}}
\end{center}
\end{figure}

In what follows, we assume that the primary energy
estimates for hybrid events will be accurate at a few percent level.
In fact, simulations show that this accuracy for 50\% of events improves from
9.5\% at 10$^{18}$eV to 2.5\% at 10$^{20}$eV \cite{AugerRep}.  
The data set used to implement and test the
methods described in the following sections consists of 8000
vertical cascades produced by particles with the fixed energy of 1
EeV, simulated using the CORSIKA program (version 6.004) \cite{Heck} 
with the QGSJet98 \cite{qgsjet} hadronic interaction model.
Simulations were performed at the Lyon Computer Centre.
The primary nuclei were $P$, $He$, $O$ and $Fe$, each of them initiating 2000
cascades. The CORSIKA output provides the number of charged
particles at atmospheric depths sampled with 5 g cm$^{-2}$ intervals.\\
We clipped the data at a depth of 200 g
cm$^{-2}$, since the FD detection threshold does not allow to
detect the weak signals at the beginning of the cascade
development. The maximum atmospheric depth was set at 870
g cm$^{-2}$, roughly corresponding to the level
of the Pierre Auger Observatory. In Figure \ref{fig:fcc0}
we show a subsample of 50 cascades for each primary.
\section{Fit of the Cascade Curve (FCC)}
Many methods for the determination of the primary mass
composition are based on the fit to the distribution of a variable
sensitive to the primary mass by a set of simulated distributions
obtained for pure primaries and for their weighted combinations. 
Values of partial amplitudes obtained as a
result of such a fit are then used as a measure of the abundance of
different nuclei in the observed mass composition, thus giving the
{\it mean mass composition} without, however, identifying the
origin of each individual cascade.
The first requirement is to have a large number of
experimental cascades in order to build, with sufficient statistical
accuracy, mean cascades for different bins of energy and
zenith angles and to derive the standard deviations of the particle
number at each atmospheric depth (we call it {\it mean trial
cascade}). The shape of the mean trial cascade reflects the
primary mass composition and should be fitted with a properly
weighted combination of a few template cascades derived from
simulations for different primary nuclei made with a high statistics
in order to eliminate the disturbing effect of fluctuations. Any
type of fitting algorithm can be used to derive the best fit
amplitudes which in turn provide the measure of the abundance of
their parent nuclei in the observed cosmic ray flux.
\subsection{Input data and procedure}
We divided the available statistics of 8000 simulated cascades 
into two parts, containing 4000 (4$\times$1000) and 4000 
(4$\times$1000) cascades respectively. 
The first part was used for the formation of the mean trial cascade. 
In order to put ourselves in the worse possible
condition (minimum variance) we assumed a {\bf uniform} primary
mass composition where the abundance of each constituent was 0.25
(1000 cascades for each primary nucleus). The standard
deviation of the number of particles for the mean trial cascade
was obtained by comparing all individual 4000 cascades with their
mean.\\
Other 4x1000 cascades were used to produce the 4
mean constituent cascades, which we call {\it mean test cascades}.
The fit of the mean trial cascade was then made in the range of
atmospheric depths $200\ g cm^{-2} < X < 870\ g cm^{-2}$ using the
MINUIT code.
\begin{figure}[t]
\begin{center}
\includegraphics[height=12cm,angle=0]{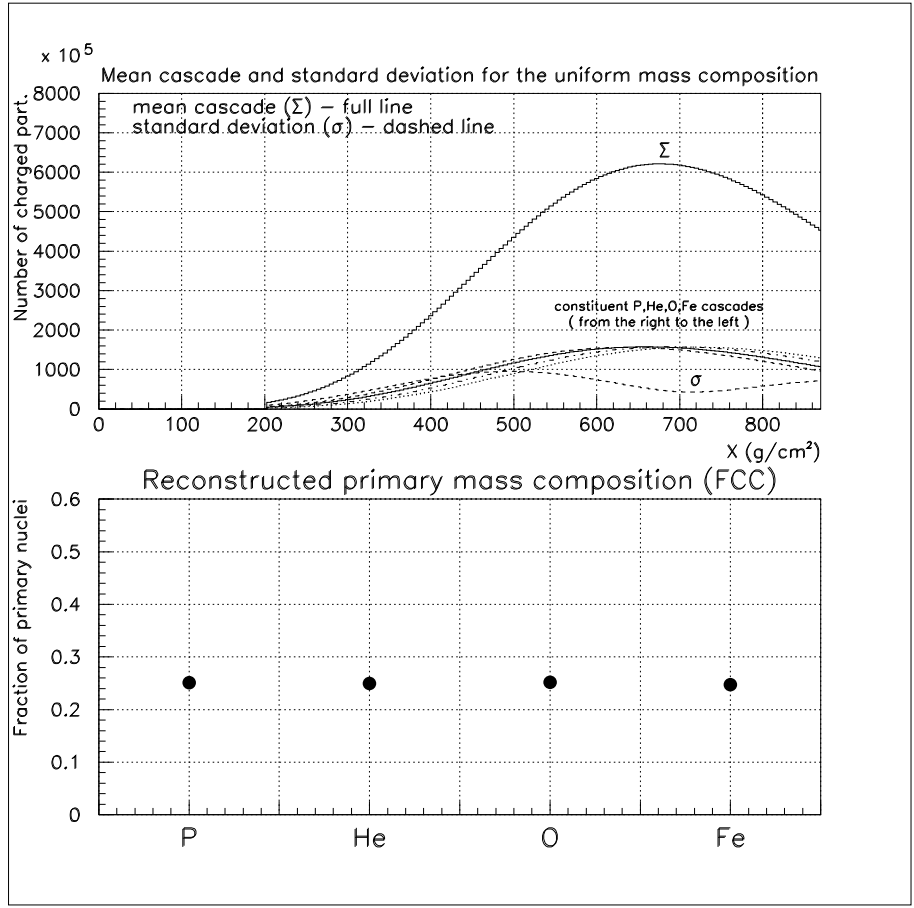}
\caption{\footnotesize Upper part: mean vertical cascade curve
($\Sigma$ - full line) and its standard deviation ($\sigma$ -
dashed line) for the uniform primary mass composition at the
energy of 1 EeV. The constituent cascade curves weighted by 0.25
for cascades from P, He, O and Fe nuclei are also shown for
comparison. Lower part: the best fit abundances of the primary
nuclei obtained by the fitting the mean cascade curve with the
MINUIT code. Errors of the obtained abundances are smaller than the size of 
the symbols. \label{fig:fcc1}}
\end{center}
\end{figure}
\subsection{Results}
The mean trial cascade curve and its standard deviation
are shown in Figure \ref{fig:fcc1} (upper panel). 
The constituent cascades taken with the weight of 0.25 are also shown. 
As it can be seen, differences between contributions of the various
constituent nuclei are small, thus making troublesome to derive their
abundance accurately, expecially in view of the
 relatively large fluctuations, both intrinsic for cascades from each
constituent nucleus, and due to the difference in the longitudinal
development between cascades from different nuclei. Therefore in order to 
derive an accurate mean mass composition with FCC a large statistics
of cascade events must be collected. \\
The result of the fit of the mean trial cascade with a set of constituent
cascades obtained using the MINUIT code is shown in Figure \ref{fig:fcc1} 
(lower panel) with its parabolic errors. Since we are interested in the mean 
mass composition we used as input errors in MINUIT not the standard 
deviations shown in Figure 2, but the errors of the mean cascade, 
which are 63 times ($\sqrt(4000)$) smaller.
The profiles of the constituent cascades were assumed to be known precisely. 
It is apparent from the Figure 2 that MINUIT reconstructs the primary mass 
composition correctly and make errors of the obtained primary abundances small.
\section{Multiparametric Topological Analysis (MTA)}
\begin{figure}[htb]
\begin{center}
\includegraphics[width=7.5cm]{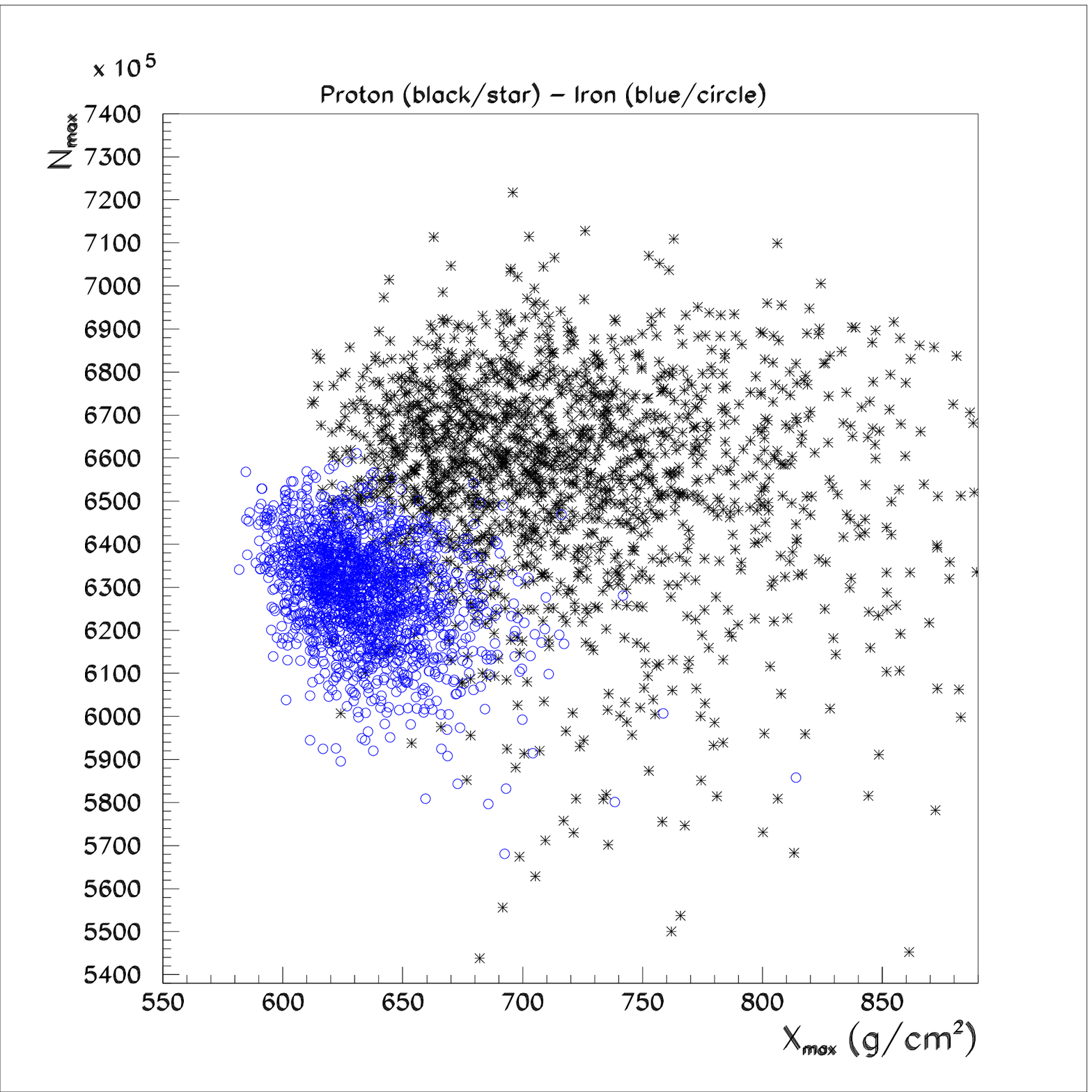}
\includegraphics[width=7.5cm]{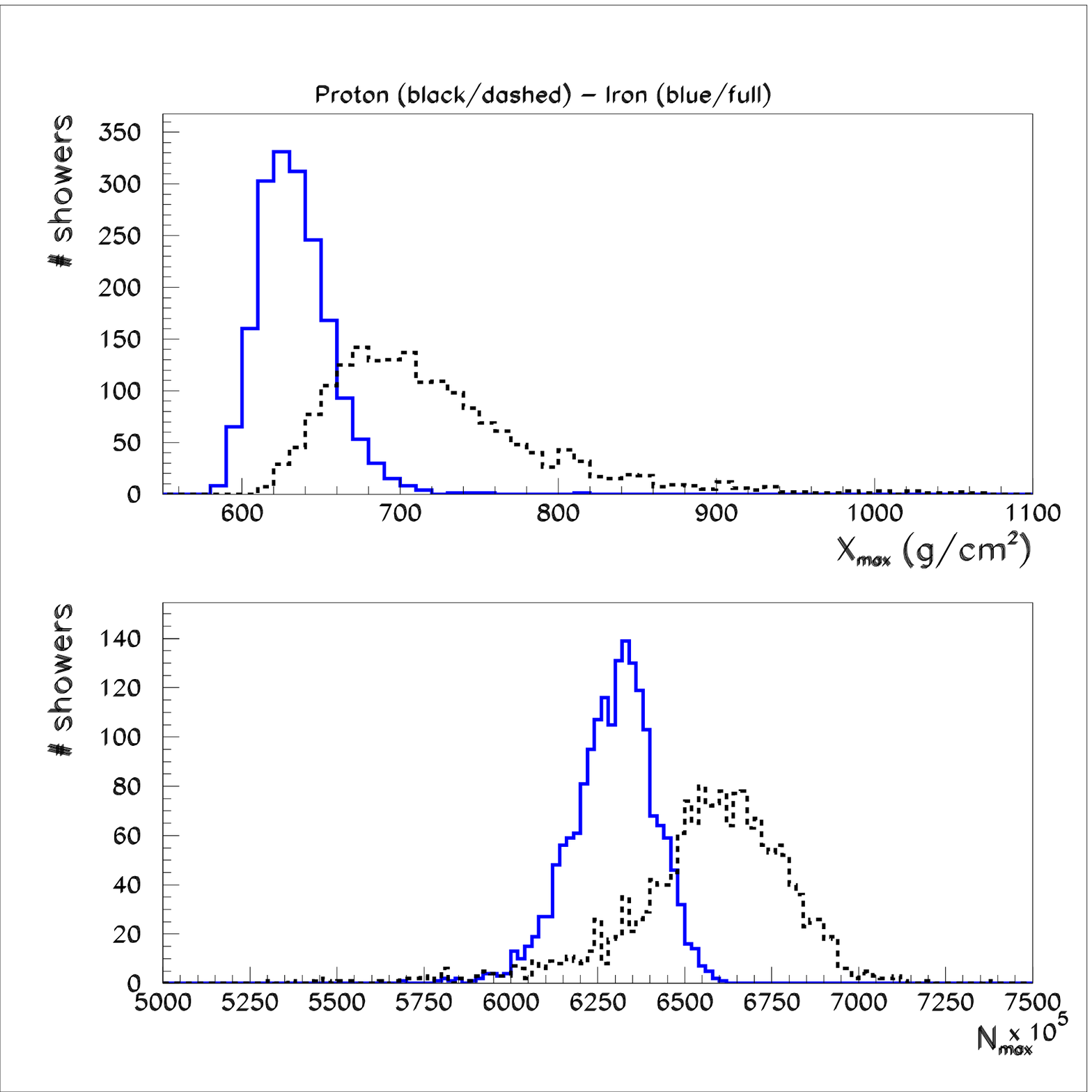}
\caption{\footnotesize a) N$_{max}$ vs. X$_{max}$ scatter plot for
proton (star) and iron (circle) induced showers.
b) Projected distributions of the scatter plot
in Figure \ref{fig:mta1} for protons (dashed line) and iron nuclei (full line)}
\label{fig:mta1}
\end{center}
\end{figure}

\begin{figure}[t]
\begin{center}
\includegraphics[height=12cm,angle=0]{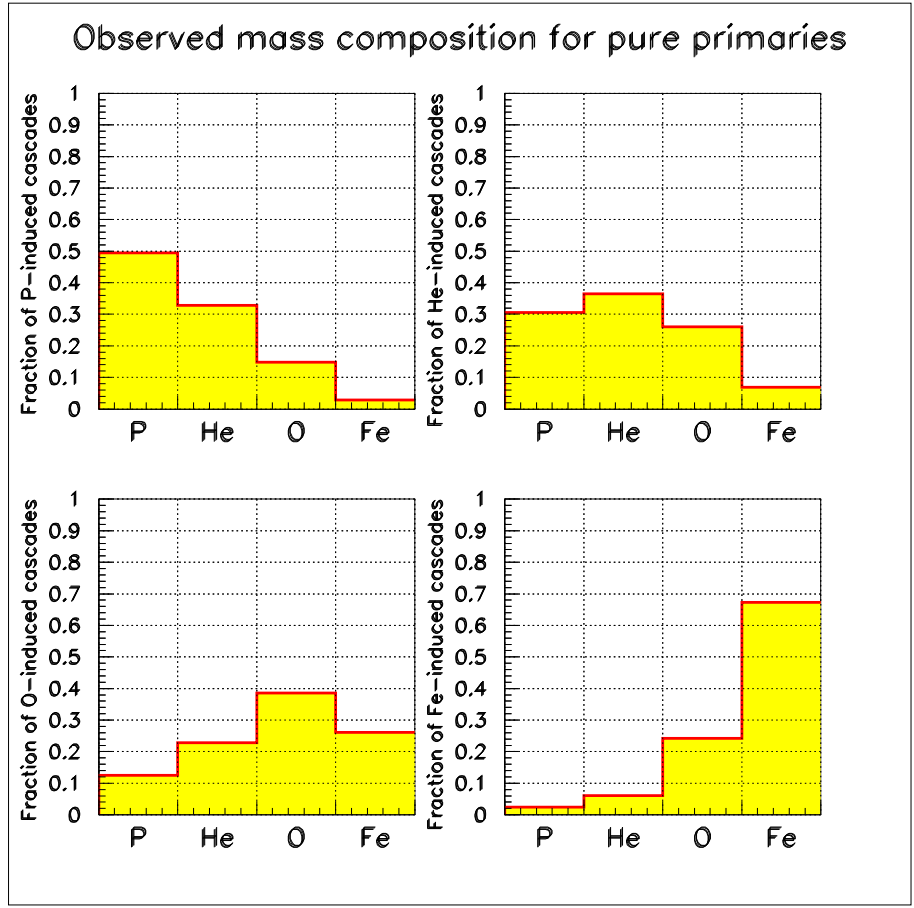}
\caption{\footnotesize Application of the MTA method to the N$_{max}$ -
X$_{max}$ scatter plot - mean probability $P_{ij}$ for the cascades induced 
by {\em i} nucleus: P, He, O and Fe (specified in headers) being identified 
as induced by {\em j} nucleus, indicated at abscissa \label{fig:mta3}.}
\end{center}
\end{figure}

The MTA method \cite{cris}, when applied to FD data, relies on a
topological analysis of correlations between the most significant
parameters of the shower development in the atmosphere.
In principle this method could be used also with a greater number of 
parameters, however, in this paper we restrict ourself to the simple case of 
two parameters only: 
X$_{max}$ (the atmospheric depth of the shower maximum) and N$_{max}$ 
(the number of charged particles at the depth of X$_{max}$). 
A scatter plot of these two parameters has been built using the
$4 \times 2000$ showers. Figure \ref{fig:mta1}
shows the scatter plot for proton-only and iron-only induced showers
as well as their projected distributions.
It can be seen that the populations arising from the two nuclei are
quite well separated in the X$_{max}$ parameter and less in
the N$_{max}$ parameter. The MTA method consists in dividing the
scatter plot in cells whose dimensions are defined by the
accuracy with which the parameters can be measured. In our simulations the 
value of 20 g/cm$^2$ has been used as the width of X$_{max}$
bin, while 5$\times 10^6$ was assumed for the width of N$_{max}$ bin. 
In each cell we can define the total number
of showers N$_{tot}^{i}$, as the sum of N$_P^{i}$, N$_{He}^{i}$,
N$_O^{i}$ and N$_{Fe}^{i}$ showers induced by P, He, O and Fe
respectively, and then derive the associated frequencies:
p$_P^{i}$=N$_P^{i}$/N$_{tot}^{i}$, p$_{He}^{i}$=N$_{He}^{i}$/N$_{tot}^{i}$, 
p$_O^{i}$=N$_O^{i}$/N$_{tot}^{i}$ and
p$_{Fe}^{i}$=N$_{Fe}^{i}$/N$_{tot}^{i}$ which can be interpreted
as the probability for a real shower falling into the $i^{th}$ cell to
be initiated by proton, helium, oxygen or iron primary nuclei.
In other words, in the case of an experimental data set of $N_{exp}$
showers, it may be seen as composed by a mixture of N$_{exp}$ x
p$_P$ proton showers, N$_{exp}$ x p$_{He}$ helium showers,
N$_{exp}$ x p$_O$ oxygen showers and N$_{exp}$ x p$_{Fe}$ iron
induced showers, where p$_A = \Sigma^i$p$_A^i/$N$_{exp}$. \\
In order to generate the $X_{max} - N_{max}$ scatter plot and produce the 
relevant matrix of cells we used a set of 4$\times$1000 simulated showers. 
Then we used another subset of 4$\times$600 showers to determine the 
probabilities $P_{ij}$. For each individual shower $i$ in a given subset the 
partial probabilities p$_P^i$, p$_{He}^i$, 
p$_O^i$ and p$_{Fe}^i$ have been read from the relevant cell {\em i}. 
The sum of such probabilities over the entire subset of 600 showers permits
to estimate the probability for a shower of a given nature $A$ to be
identified as a shower generated by $P$, $He$, $O$ or $Fe$ primary particle.
This probability is shown in Figure \ref{fig:mta3}. 
One can see that the method attributes the highest probability to the correct 
nuclei.
The application of MTA for the determination of the mean primary mass 
composition will be described later and compared with the results of other 
methods. 
\section{The Minimum Momentum Method (MMM)}
\begin{figure}[t]
\begin{center}
\includegraphics[height=14cm,angle=0]{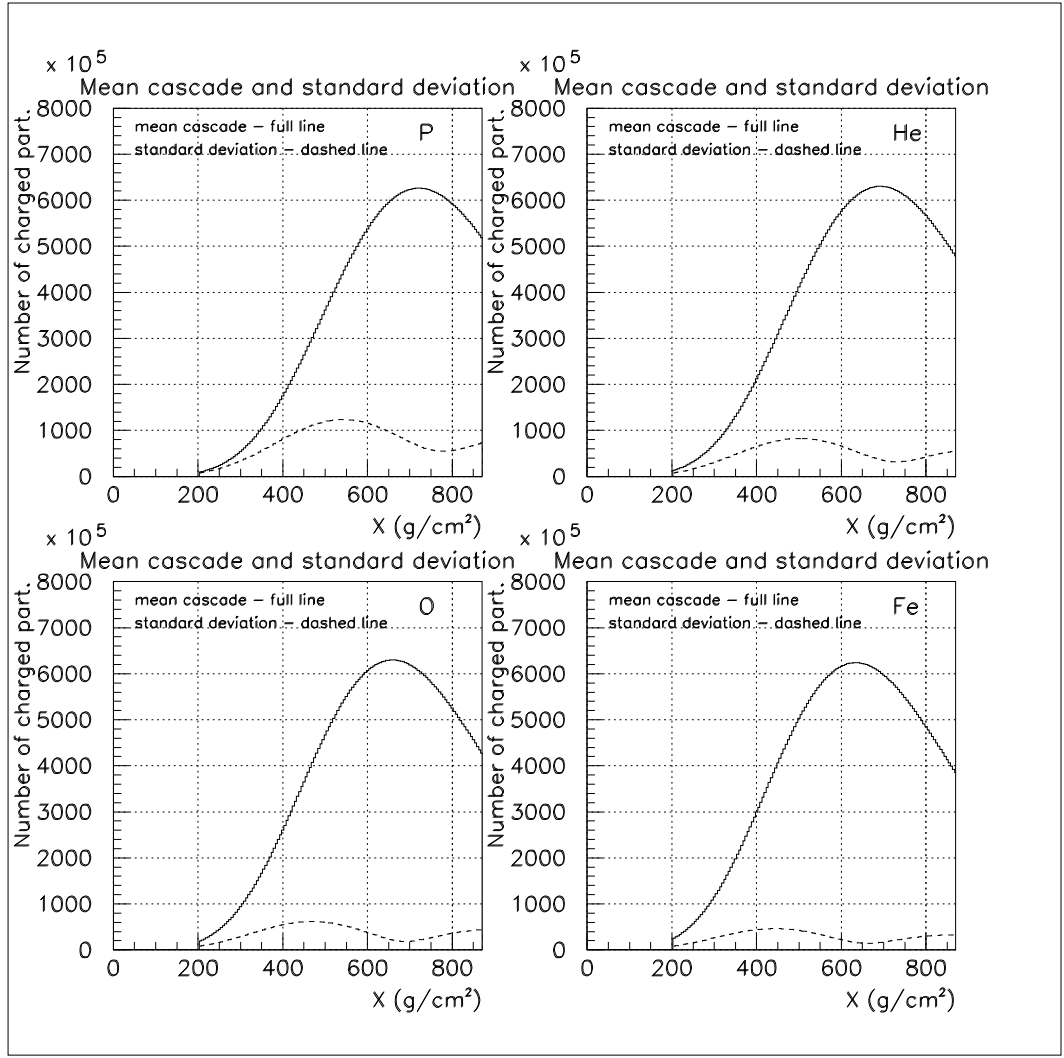}
\caption{\footnotesize The mean cascade (full line) and the
standard deviation of its particle number (dashed line) for the
{\em vertical} cascades initiated by primary protons ($P$), helium
($He$), oxygen ($O$) and iron ($Fe$) nuclei (indicated in the
upper right corner of the graphs) with the energy of 1 EeV. The
abscissa is the atmospheric depth in g cm$^{-2}$, the ordinate is
the number of charged particles.\label{fig:mmm1}}
\end{center}
\end{figure}

\begin{figure}[htbp]
\begin{center}
\includegraphics[height=16cm,angle=0]{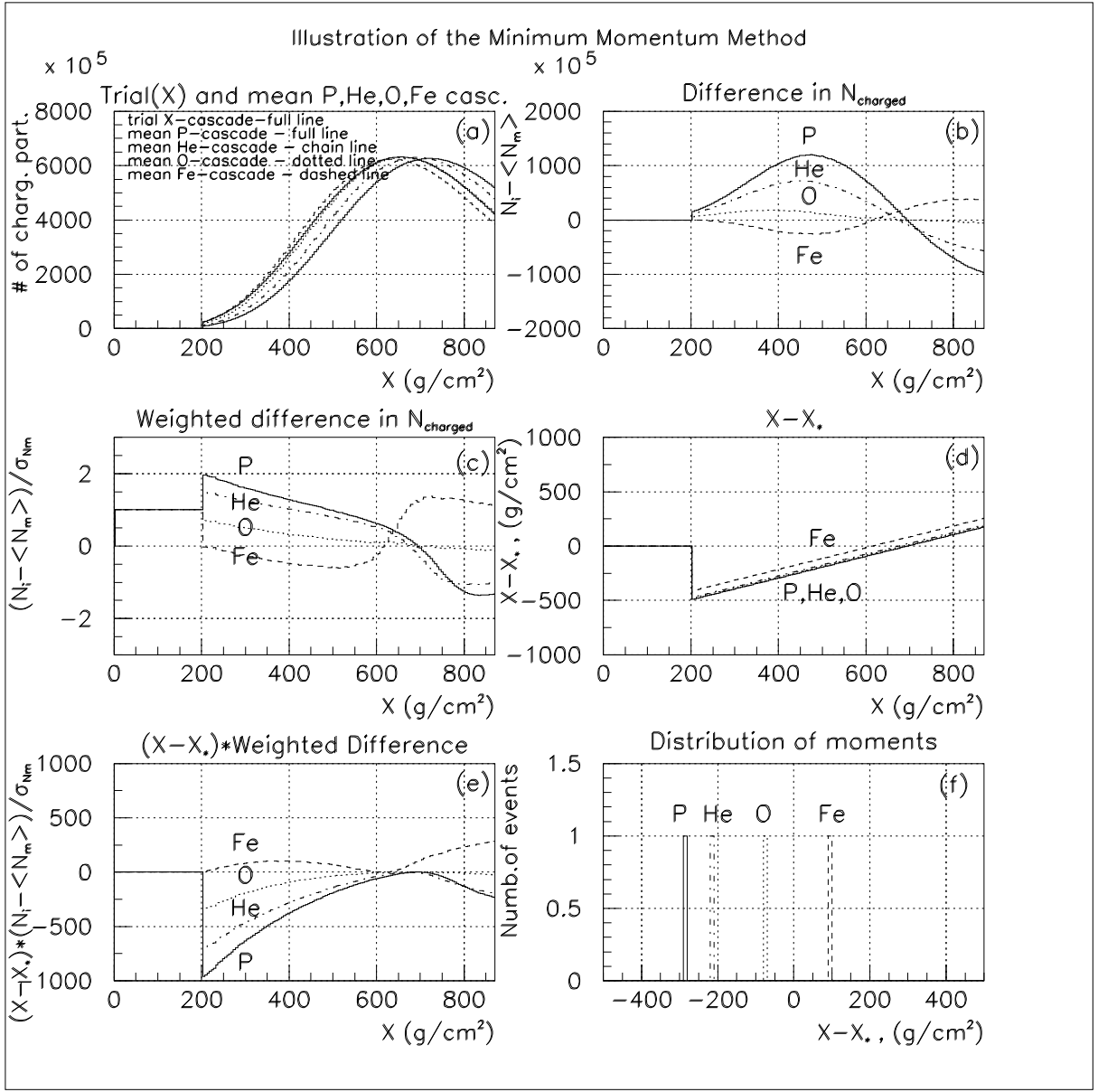}
\caption{\footnotesize The Minimum Momentum Method. Panel (a): the
trial cascade (left solid line) and the mean test cascades,
induced by primary $P$ (right full line), $He$ (dash-dot line),
$O$ (dotted line) and $Fe$ (dashed line) nuclei, with which the
trial cascade is compared. Panel (b): difference in the particle
number between the trial and test cascades. The notations here and
in the subsequent graphs are the same as in panel (a). Panel (c): the
difference in particle numbers weighted with the standard
deviations $\sigma^{N_m}_i$ shown in Figure \ref{fig:mmm1}. Panel
(d): the atmospheric depth rescaled with the depth $X^\ast_{lm}$
corresponding to the crossing point of the trial and test
cascades. Panel (e): the rescaled atmospheric depth multiplied by
the weighted difference in the particle number. Panel (f): the first
momentum of the weighted difference, obtained just by an
integration of the functions shown in panel (e). The minimum of the
absolute values of these four momenta defines the origin of the
trial cascade (as an example we show an Oxygen nucleus).\label{fig:mmm2}}
\end{center}
\end{figure}

\begin{figure}[t]
\begin{center}
\includegraphics[width=11cm,angle=0]{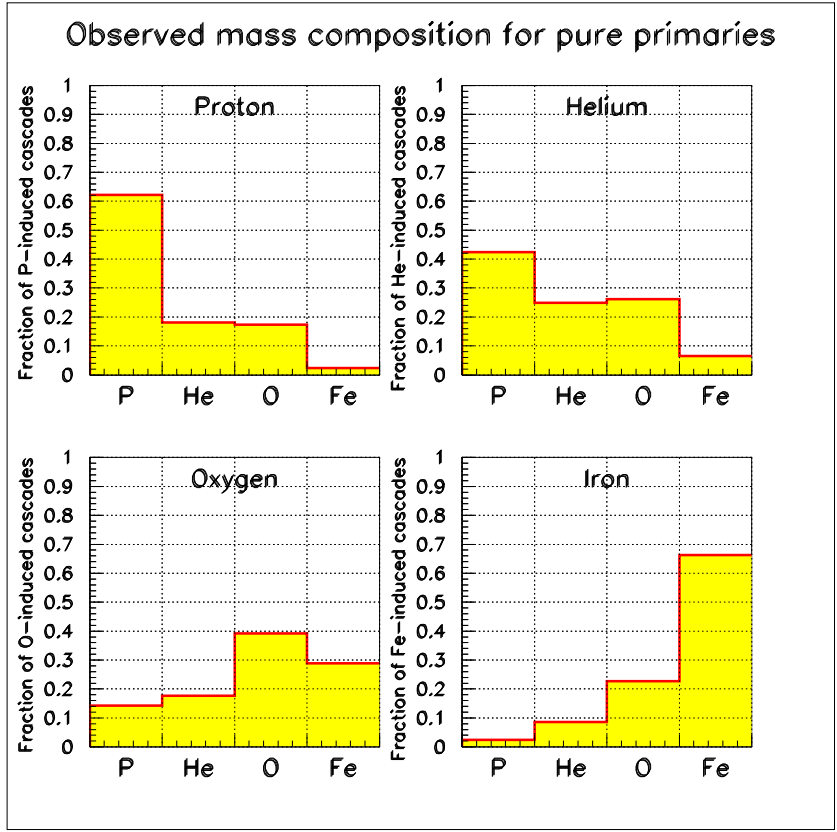}
\caption{\footnotesize The mean probability for the primary P, He, O and Fe 
nuclei (specified in the headers of the graphs) to be identified as P, He, O 
and Fe by MMM.  In the MMM each trial $i$ cascade has been associated 
with {\em just one} class of test $j$ cascades.
\label{fig:mmm4}}
\end{center} 
\end{figure}

Besides various methods of the analysis of the mean mass composition outlined 
above there is also the possibility to identify the mass of the primary 
particle for each individual cascade and therefore to determine the observed 
mass composition. The idea behind this method originated from the well
known KNN ('K Nearest Neighbours') method. In the MMM method \cite{mmm} as a
measure of the closeness  between trial ({\em l}) and test ({\em m}) 
cascades we decided to use the distance $D_{lm}$,
which incorporates all the available information. It takes into
account:
\noindent (i) the longitudinal development of cascades, i.e. the
function $N_{ch}(X)$, where $N_{ch}$ is the number of charged particle
at the atmospheric depth $X$;
\noindent (ii) the fluctuations of the cascade development;
\noindent (iii) the mutual position of the compared cascade
curves, i.e. whether the test cascade develops at greater or at
lower atmospheric depths with respect to the trial cascade.\\
This has been achieved by introducing the following definition of
distance:
\begin{equation}
 D_{lm}=abs[\Sigma_i(X_i-X^\ast_{lm})\frac{N^l_i-\langle N^m_i \rangle}{\sigma^{N_m}_i}]= abs(M_{lm})
\label{eq:dist}
\end{equation}
\noindent Here $N^l_i$ is the number of charged particles in the trial cascade at the 
depth of $X_i$, $\langle N^m_i \rangle$ is the number of charged particles at the same 
depth $X_i$ in the mean
cascade initiated by the primary nucleus $m$, where $m$ stands for $P, He, O, Fe$.
$\sigma^{N_m}_i$ is the standard deviation of $N^m_i$ at the depth $X_i$ and $M_{lm}$ 
is the first momentum of the weighted difference. The mean cascades and the standard 
deviations of their charged particle numbers as a function of the atmospheric depth
$X$ are shown in Figure \ref{fig:mmm1}.
Interestingly, the minimum fluctuations is not at
the mean depth of the maximum development $X_{max}$ but slightly
shifted to the larger depths. This is consequence of the fact
that besides the ordinary fluctuations of the particle number
there are also fluctuations in the position of the first
interaction point (~starting points of the cascade development~).
Also remarkable is the fact that the standard deviations do not decrease with
increasing primary mass $A$ as $\frac{1}{\sqrt A}$ as it is expected 
for the superposition model. For instance,
the fluctuations in iron induced cascades are smaller only by a
factor of 2.6 compared with those of proton induced cascades
instead of $\sqrt{56} \approx 7.5$. This result confirms the
non-validity of the superposition model often used for the
estimates.\\
The fluctuations play an important role since, for instance, in
the case of an equal difference in the particle numbers of the
trial and mean test cascades, the method favors the mean cascade
with the smallest fluctuations.\\
In order to include the information about the relative location of
the trial and the mean test cascade in the atmosphere, we used the
term $(X_i - X^\ast_{lm})$. Here $X^\ast_{lm}$ is the depth at
which two cascade curves cross (Figure \ref{fig:mmm2}).
In panel $(a)$ of the figure, the trial cascade which is shown by
the left solid line, is compared with four test cascades. They are
the mean cascades induced by P, He, O and Fe nuclei, as those
shown in Figure \ref{fig:mmm1}. The difference in the particle
numbers between the trial and test cascades is shown in panel $(b)$.\\
It may be seen that:\\
(i) the cascades, which are 'to the right' of the trial cascade
($P$, $He$ and $O$) give a different difference profile with
respect to that 'on the left' ($Fe$);\\
(ii) the crossing point $X^\ast_{lm}$ moves to the left from $P$
to $Fe$.\\
The weighted difference in particle numbers $\frac{N_l - \langle
N_m \rangle}{\sigma^{N_m}_i}$ is shown in panel $(c)$. Since all
standard deviations are positive, the weighted differences
preserve the same sign as the original differences, i.e. they are
positive below the crossing point $X^\ast_{lm}$ for the 'right'
cascades and negative for the 'left' ones. Above the crossing
point they change sign.\\
If we simply integrate these curves, the positive and negative
parts partly compensate each other and the sensitivity of such
integral to the primary mass is reduced. This is why we decided to
make the integration for the function which is the product of the
weighted difference and the first momentum rescaled to $0$ at the
crossing point: $X - X^\ast_{lm}$. This rescaled momentum is shown
in panel $(d)$. It also changes its sign at the crossing point. When
we multiply these functions for 'right' ($P$, $He$, $O$) cascades
the product is negative in the whole range of atmospheric depths,
both below and above the crossing point. The same is true for the
'left' ($Fe$) cascades, but in this case  the product is positive. The
product functions $(X_i-X^\ast_{lm})\frac{N^l_i-\langle N^m_i
\rangle}{\sigma^{N_m}_i}$ are shown in panel $(e)$. Different signs
of the functions for $P$, $He$, $O$ and $Fe$ induced cascades are
clearly seen. When we integrate these functions, we obtain the
values of the first momentum which have different signs for
'right' and 'left' cascades (panel $(f)$). In this way we not only
increase the separation between the trial and different test
cascades, but we also determine whether the test cascades have
earlier or later development in the atmosphere with respect to the
trial cascade.\\
The last problem to be addressed is how to use these momenta for
the determination of the parent nucleus. We define it to be the
test nucleus which gives the cascade closest to the trial cascade
in terms of the distance $(1)$. Therefore we first determine the
distances as the absolute value of the momentum $(1)$ and then
find their minimum $D_{min} = min(D_{lm})$. The nucleus $'m'$
which gives this minimum is defined as the parent nucleus for the
trial cascade. Hence we call this method as the {\it Minimum
Momentum Method (MMM)}.\\
The probabilities $P_{ij}$ for the cascades induced 
by primary {\em i} nuclei (P, He, O and Fe) being identified as induced by 
{\em j} nuclei are shown in Figure \ref{fig:mmm4}.
This figure demonstrates that the MMM method works reasonably well in 
distinguishing light nuclei (P and He) from heavy nuclei (O and Fe), but 
has not sufficient accuracy to identify separately He and O nuclei.
The application of MMM for the determination of the mean primary mass 
composition will be described and compared with the results of other 
methods below. 
\section{The Neural Net Analysis (NNA)}
Neural nets are known to be among the best tools to tackle classification and
pattern recognition problems. A Neural Net (hereafter NN) is
usually structured into an input layer of neurons, one or more
hidden layers and one output layer; neurons belonging to adjacent
layers are usually fully connected and the various types and
architectures are identified both by the different topologies
adopted for the connections as well as by the choice of the activation
function. The values of the functions associated with the
connections are called ``weights'' and the whole game of NN's is
in the fact that, in order for the network to yield appropriate
outputs for given inputs, the weights must be set to a suitable
combination of values \cite{Bishop}. The way this is obtained
leads to the first important difference among modes of operations,
namely between ``supervised'' and ``unsupervised'' methods.\\
In supervised methods, the network learns by examples and
therefore, the user needs to know the correct output value for a
fair subsample of the input data. This set needs to be divided
into other three subsets named, respectively, {\it Training}, {\it
Validation} and {\it Test} (T/V/T) sets. The first subset is used to fine
tune the weights, the second one to check whether the network has
achieved an acceptable generalization capability and, finally, the
third subset is used to evaluate the performances and the classification 
errors.\\
In unsupervised methods, instead, input data are clustered on
the basis of their statistical properties only. Whether the
obtained clusters are or are not significant to a specific problem
and which meaning has to be attributed to a given cluster, is not
obvious and requires an additional phase called
``labeling''. The labeling requires that the user knows the
characteristics of a small sample of input vectors (labeled set).\\
We tested both supervised and unsupervised methods.\\
The unsupervised experiments were performed using a Self
Organizing Map (SOM) \cite{Kohonen} with 120 neurons and as
labeling set we made use of the same 8000 simulated curves already described. 
Each neuron was then attributed to a specific class
accordingly to the type of event which had activated that neuron
more times. Results may be summarised as follows: $P$ = 34\%
success rate, $He$ = 30\%, $O$ = 28\%, $Fe$=41\%.
\begin{figure}
\begin{center}
\begin{tabular}{c}
\includegraphics[height=8cm]{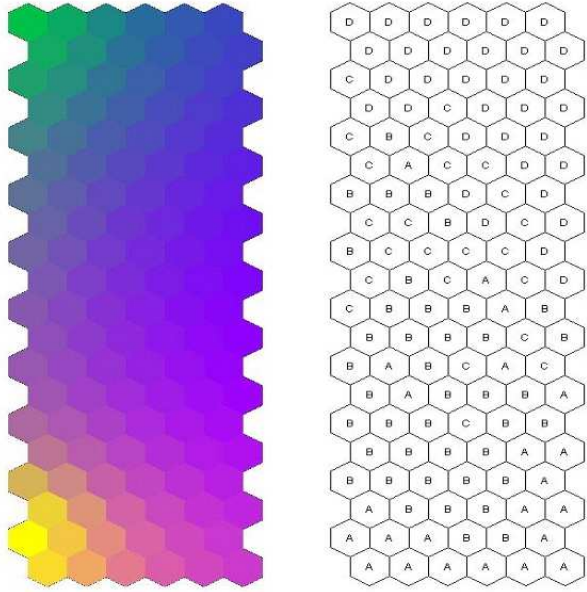}
\end{tabular}
\end{center}
 \caption{\footnotesize Left panel: SOM similarity coloring map: each hexagon
 represents a neuron and different colors denote different
 clusters. Right panel: neurons are labeled 
 (A=proton; B=Helium; C=Oxygen; D=Iron).For each neuron the class is 
determined by the type of nucleus which activetes that neuron the largest 
number of times 
\label{AUGER_unsupervised}}
\end{figure}
Supervised experiments were instead performed using a Multi Layer Perception 
(MLP) with Bayesian learning \cite{Bishop} and auxiliary sets extracted from
the above quoted simulated curves (for each primary, 1000 input for
the training set, 600 and 400 for the validation and test sets
respectively). As discussed above, the training set provided the
"a priori" knowledge, the validation set ensured that after
training the network still had enough generalization capabilities
(thus preventing overfitting, i.e. the fact that the network
learns to recognize only the data on which it was trained), the
test set is used for evaluating the performance of the network. 
To be as realistic as possible, i.e., to operate in absence of a priori 
knowledge as it would be the case for real data, the extraction of the TVT set
was made on the basis of the unsupervised clustering which allowed to identify
the most significant data subsets and to extract TVT data accordingly. 
In order for the whole procedure to be effective, all three sets need
to be statistically representative of the data to be processed: i.e.
 they need to sample homogeneously the parameter space. In
order to obtain for each input vector the probability that it
belongs to a given class, we adopted the SoftMax activation
function and the Entropy function as error estimator. Several sets of 
experiments were performed.
 \begin{figure}[t]
 \begin{center}
 \includegraphics[height=11cm]{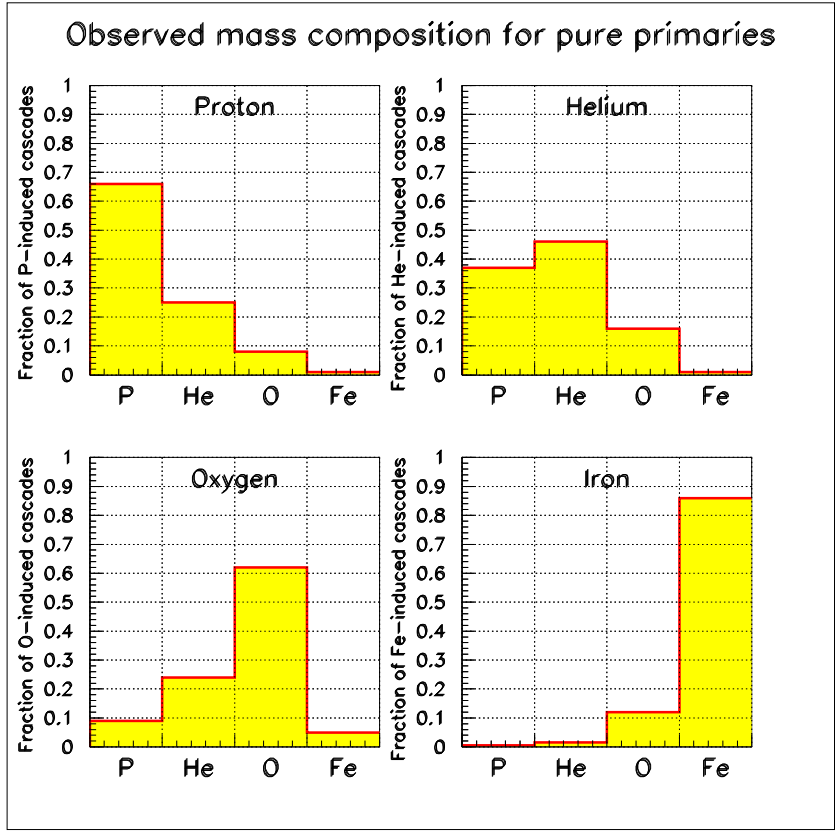}
 \end{center}
 \caption{\footnotesize Results for MLP (22 hidden neurons, SoftMax
 activation function) with conjugate gradient optimization
 algorithm. The diagrams give the distribution among the four types of primary
 particles as obtained after labeling.\label{AUGER_supervised}}
 \label{fig:AUGER_MLP}
 \end{figure}
In a first set of experiments we trained the network on the $6$
parameters resulting from the model driven fit of the
Gaisser--Hillas curves and therefore the adopted MLP consisted of
$6$ input neurons, one hidden layer with a number of
neurons variable from experiment to experiment 
and $4$ output neurons. Each output neuron corresponds to
a class (Proton, Helium, Oxygen, Iron) and provides the
probability that the input event belongs to that class.
In a second set of experiments we instead trained the net using the
entire simulated curve (174 parameters), finding no appreciable
difference regardless the adopted NN Architecture - a fact which
should not surprise since we are dealing with noiseless data and the best fit
curve is a truly good approximation to the data.
Results are shown in Figure \ref{fig:AUGER_MLP} for the case in which the net 
had 22 neurons in the hidden layer with conjugate gradient optimization
algorithm.\\
We have used different activation functions (AF) and we have seen they produce
quite different results. For instance, the Descendent Gradient AF
achieves high accuracy in disentangling the two extreme cases of
proton and Iron, but has lower performances for 
intermediate masses (He and O) which are better separated by the
Conjugate Gradient AF \cite{Bishop}.
From the above results it is apparent that even in the absence of
a fine tuning of the networks, supervised methods outperform
unsupervised ones.\\
It needs to be stressed that, once the network has been trained and frozen,
its application to new input data vectors leads to the attribution of the
input vector to one and only one of the 4 output neurons, and therefore,
each input vector is attributed to one among the four possible classes
with an error attached to the individual data point. As in the other methods,
however, errors may be estimated as statistical using the test set and the 
same formalism detailed in the following section. 
The application of NNA for the determination of the mean primary mass 
composition and the comparison of its results with results of other methods 
will be described in the next section. 
\section{\bf Determination of the primary mass composition}
\begin{figure}[hbt]
\begin{center}
\includegraphics[height=12cm,width=12cm,angle=0]{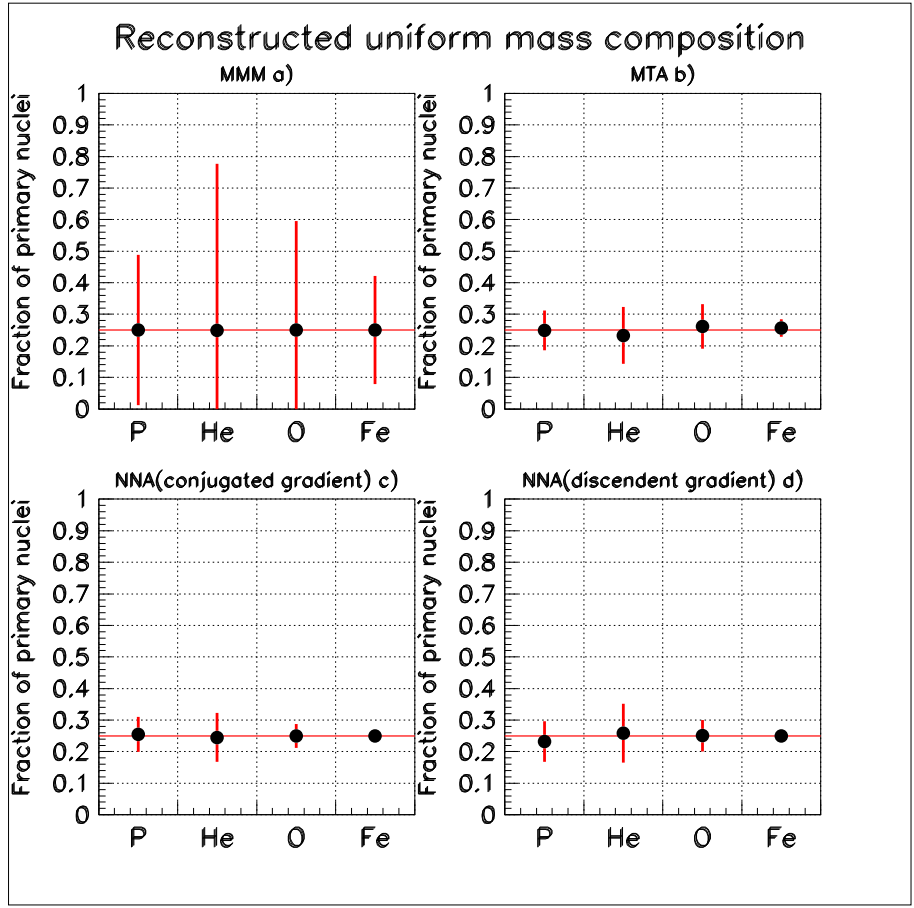}
\caption{\footnotesize The uniform mass composition 
($\Delta_P = \Delta_{He} = \Delta_O = \Delta_{Fe} = 0.25$) 
reconstructed from the observed mass composition by (a)-MMM, (b)-MTA, 
(c)-NNA with conjugated gradient optimization 
algorithm, (d)-NNA with discendent gradient optimization algorithm methods
\label{fig:unif}}
\end{center}
\end{figure}

The obtained mean probabilities $P_{ij}$ for the primary mass $i$ to be 
identified as the mass $j$ in the case of pure primary mass composition 
can be used for the reconstruction of the mixed primary mass composition as the
coefficients in the system of linear equations:
\begin{eqnarray}
\nonumber n^{\prime}_P & = & n_P \cdot P_{P \rightarrow P} + n_{He} \cdot  
P_{He \rightarrow P} + n_O \cdot P_{O \rightarrow P} + 
n_{Fe} \cdot P_{Fe \rightarrow P} \\
n^{\prime}_{He} & = & n_P \cdot P_{P \rightarrow He} + n_{He} \cdot 
P_{He \rightarrow He} + n_O \cdot P_{O \rightarrow He} + 
n_{Fe} \cdot P_{Fe \rightarrow He} \\
\nonumber n^{\prime}_O &= &n_P \cdot P_{P \rightarrow O} + n_{He} 
\cdot P_{He \rightarrow O} +  n_{He} \cdot P_{O \rightarrow O} + 
n_{He} \cdot P_{Fe \rightarrow O} \\
\nonumber n^{\prime}_{Fe} & =& n_P \cdot P_{P \rightarrow Fe} + n_{He} 
\cdot P_{He \rightarrow Fe} + n_O \cdot P_{O \rightarrow Fe} +
n_{Fe} \cdot P_{Fe \rightarrow Fe} 
\end{eqnarray}
where $n_P$, $n_{He}$, $n_O$ and $n_{Fe}$ are the true numbers, defining the 
primary mass composition in the sample 
$N=n_P+n_{He}+n_O+n_{Fe}$, which are altered to $n^\prime_P$, $n^\prime_{He}$, 
$n^\prime_O$ and $n^\prime_{Fe}$, 
due to a misclassification. \\
If we define  $\Delta_i=\frac{n_i}{N}$ we can rewrite the system (2) as:
\be
\Delta_j^\prime = \Sigma_{i=1}^4 P_{ij} \Delta_i
\ee
where $j = 1-4$, $\Delta_j^\prime$ is the observed abundance of the primary 
mass $j$, $\Delta_i$ is the true abundance of the mass $i$ in the primary 
cosmic rays with the constraints:
\be
\Sigma_{i=1}^4 \Delta_i = \Sigma_{j=1}^4 \Delta_j^\prime = 1
\ee
In order to invert the problem and to reconstruct the abundances
$\Delta_{A_i}$ in the primary mass composition from the observed
abundances $\Delta^\prime_{A_i}$ and the known probabilities $P_{ij}$, we can 
apply any method capable to solve the inverse
problem taking into account possible errors of the observed
distribution with the constraints (4).\\
In what follows, we use the MINUIT code. The observed abundances were 
simulated using another subset of 4$\times$400 cascades different from those 
used for determination of $P_{ij}$. The errors of the observed abundances 
$\Delta_A^\prime$ were derived as the errors of the mean for the total number 
of 1600 cascades. 
Probabilities $P_{ij}$ were assumed to be known precisely for the MMM and for 
NNA methods. In this case MINUIT solves the system by the least-square method, i.e. minimizing the function
\be
\chi^2 = \Sigma_{j=1}^4 \frac{(\Delta_j^\prime - \Sigma_{i=1}^4 P_{ij} 
\Delta_i)^2}{\epsilon_j^2}
\ee
where $\epsilon_j$ is the rms error of the difference in the numerator.
In the case when probabilities $P_{ij}$ are known precisely this error 
contains only errors of the observed abundance, i.e. $\epsilon_j = 
\sigma_{n_j^\prime}$. Due to the constraint (4) we applied rms errors of 
$\Delta_j^\prime$, assuming a multinomial distribution of the number of 
cascades \cite{Eadie}. Actually the multinomial distribution 
is valid not for the sum of cascades, identified as those induced by nuclei 
$j$, but separately for each constituent nucleus $i$ identified as $j$, and 
therefore the error depends on the primary mass composition. 
The relevant expression for the error in this case is
\be
\epsilon_j^2 = \sigma_{\Delta_j^\prime}^2 = \frac{\Sigma_{i=1}^4 
\Delta_i P_{ij} (1-P_{ij})}{N}
\ee   
where $N$ is the total number of 4x400 cascades. By this way we introduce a 
non-linearity into the minimized function $\chi^2$, which now becomes a 
non-linear function of unknown variables $\Delta_i$. 
However, MINUIT is able to 
solve this problem even in this case. \footnote{If the probabilities $P_{ij}$ 
have the maximum for the correct identification $P_{ii}$, then the 
distribution of observed abundancies $\Delta_j^\prime$ is similar to the 
distribution of true abundancies $\Delta_i$. In this case it is possible 
to use an approximate expression $\epsilon_j^2 = \frac{\Delta_j^\prime 
(1-\Delta_j^\prime)}{N}$, 
which gives only slightly larger errors than the precise expression (6).}\\
The results of the solution for the uniform primary mass composition with all 
abundances $\Delta_A= \frac{n_A}{N}$ = 0.25 reconstructed by MMM, MTA and NNA 
methods is shown in Figure \ref{fig:unif}. 

\begin{figure}[hbt]
\begin{center}
\includegraphics[height=12.5cm,width=12.5cm,angle=0]{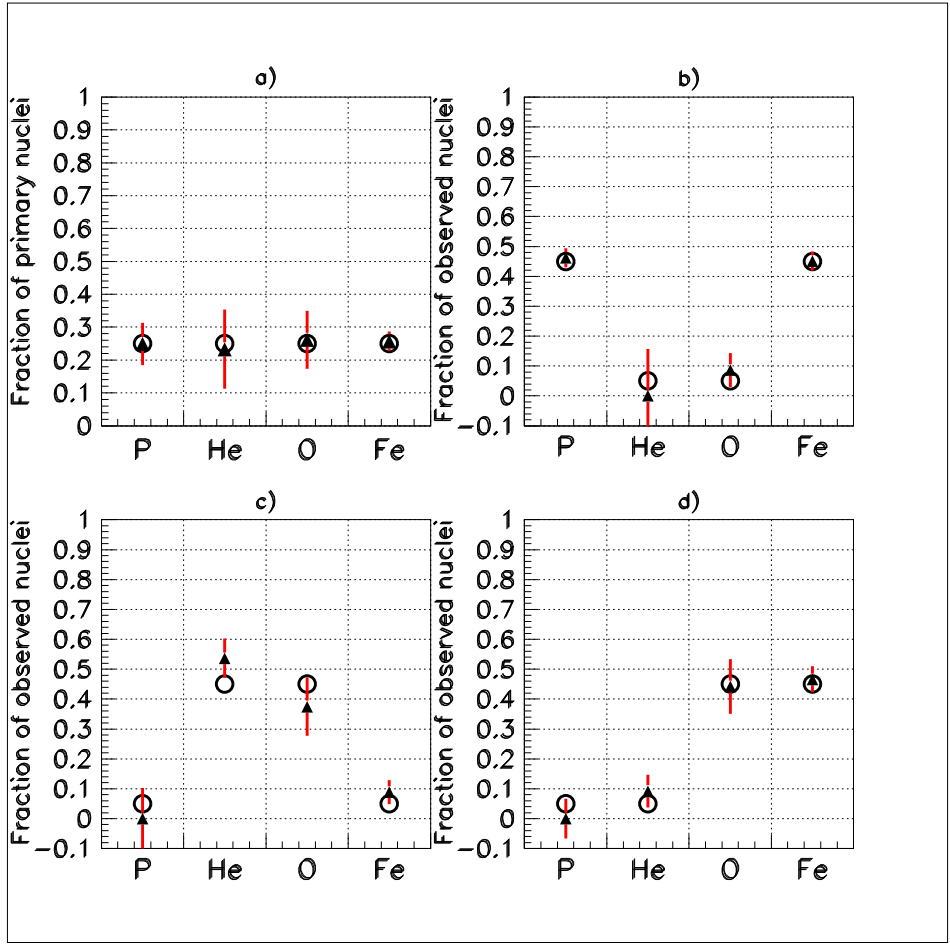}
\caption{\footnotesize The same as in Figure \ref{fig:unif} for MTA method, 
taking into account the errors of probabilities, in 
the case of uniform (a) and non-uniform (b,c,d) primary mass composition for 
three mass compositions. The empty circles are the true abundancies and the 
full triangles are the reconstructed ones.  
\label{fig:nonunif}}
\end{center}
\end{figure}

In the case of MTA it is possible to calculate the errors on the probabilities 
taking into account that $P_{ij}$ is the
{\em mean} value of probabilities $P_{ij}^m$ in the bins of the grid hit by 
the cascades $i$, averaged over all these $M$ cascades:
\be
P_{ij} = \Sigma_{m=1}^M P_{ij}^m/M
\ee
Since all $P_{ij}^m$ are independent of each other, we can write
\be
\sigma_{P_{ij}} = \sqrt{\Sigma_{m=1}^M \sigma_{P_{ij}^m}^2}/M
\ee
As an example we examine the case when $i = 1$ and $j = 1,2,3,4$. 
For $P_{1j}^m$ we can write
\be
P_{1j}^m = \frac{n_1^m}{\Sigma_j^4 n_j^m}
\ee
or 
\be
P_{1j}^m = \frac{1}{1+\Sigma_{j=2}^4 (n_j^m/n_1^m)}
\ee
Mind that $n_j^m$ in the bins are independent of each other. Therefore we can 
write
\be
\sigma_{P_{1j}^m}^2 = (P_{1j}^m)^4 \Sigma_{j=2}^4 \sigma_{(\frac{n_j^m}
{n_1^m})}^2
\ee
The distribution of the number of cascades $n_j^m$ in the bin is also 
multinomial, but for the large statistics used for the formation of the grid 
and a small bin size we can approximate it by the Poissonian distribution and 
write:
\be
\sigma_{(\frac{n_j^m}{n_1^m})}^2 = (\frac{n_j^m}{n_1^m})^2 \lbrack 
(\frac{\sigma_{n_j^m}}{n_j^m})^2 + (\frac{\sigma_{n_1^m}}{n_1^m})^2 \rbrack = 
(\frac{n_j^m}{n_1^m})^2 (\frac{1}{n_j^m} + \frac{1}{n_1^m})  
\medskip
\ee
Combining these equations we obtain the errors of the coefficients $P_{1j}$. 
The same should be done for $i = 2,3,4$. \\
At the end to account for the inaccuracy of $P_{ij}$, in the case of non-zero 
values of $\sigma_{P_{ij}}$ we suggest to modify the denominator 
$\epsilon_j^2$ in the minimized function (5), including these errors into 
it as
\be
\epsilon_j^2 = \sigma_{\Delta_j^\prime}^2 + \Sigma_{i=1}^4 (\sigma_{P_{ij}} 
\Delta_i)^2 = \Sigma_{i=1}^4 \lbrack \frac{\Delta_i P_{ij} (1-P_{ij})}{N} + 
(\sigma_{P_{ij}} \Delta_i)^2 \rbrack
\ee
An additional non-linearity introduced by accounting for these errors 
$\sigma_{P_{ij}}$, does not prevent MINUIT to find the solution. \\
The case of uniform primary mass composition taking into account also the
errors on the probabilities together a few examples of the non-uniform mass 
composition in the MTA method are shown in the Figure \ref{fig:nonunif}. 
The errors of the observed abundances $\Delta_A^\prime$ were derived as the 
errors of the mean for the total number of 800 cascades (360 for P, 40 for 
He, 40 for O and 360 for Fe in the first plot of Figure \ref{fig:nonunif}, 
etc.). 
Comparing Figure \ref{fig:unif}b and Figure \ref{fig:nonunif}a it can be seen 
that the errors on the probabilities give only a slight change in the errors
of mass composition. This is due to the fact the relevant errors come from
the ${\Delta_j^\prime}$ which depend on the statistics. 

\section{Discussion}

\subsection{Comparison of the methods}

The comparison of the methods indicates that FCC gives
better accuracy in the reconstruction of the mean mass
composition than other methods. MTA and NNA give comparable accuracy. 
In principle MTA could be used 
also for an individual identification of the cascade origin like 
MMM by assigning thresholds to the probabilities $P_{ij}$. 
For example, the cascade {\em i} can be associated with primary nucleus 
{\em j} if $P_{ij}$ is the maximum value. It is easy to do if  
the number of test cascades in the MTA cell is large enough to have a 
negligible probability for an {\em equal} number of test cascades of 
different groups in the cell.
Otherwise there would be many events in which the studied cascade hits a 
cell with an equal number of test cascades of different groups and an 
individual identification becomes uncertain. \\
In the MMM method the achieved mass resolution is not high enough and 
although the identification is satisfactory for P and Fe, it is not good for 
He and O. Though MMM uses the whole available information on the cascade 
profile, it is apparent that the approach which converts all this information 
into a single distance parameter is not very efficient. \\
Certainly some other 
forms of the one-dimensional distance $D_{lm}$ can be proposed for MMM like
\begin{equation}
D_{lm} = \Sigma_i[abs\frac{N^l_i-\langle N^m_i \rangle}{\sigma^{N_m}_i}]
\end{equation}
or higher odd momenta
\begin{equation}
D_{lm}=abs[\Sigma_i(X_i-X^\ast_{lm})^n \frac{N^l_i-\langle N^m_i \rangle}{\sigma^{N_m}_i}]
\end{equation}
where $n \geq 3$ should certainly be an odd number in order not to
lose the information about the sign of the charged particle difference.
However in view of the large overlap between the longitudinal
profiles of cascades from different primary nuclei, specifically
between P and He, it is unlikely that any approach with
a single one-dimensional measure of the distance would give
substantially better results.\\
It is seen in Figure \ref{fig:unif} that the NNA method, which does not make a 
reduction of the available information for the identification of 
{\em individual} cascades gives the better accuracy than MMM.

\subsection{Application to experimental cascades}

It has to be stressed that all the results
outlined above are biased by the fact that we are dealing with the highly 
unrealistic case of simulated 'noiseless' data. 
More realistic testing has to be
performed using data which take into account the varying primary
energy, and inclination angle, the instrumental signature (noise)
and the various sources of errors (uneven and incomplete sampling,
etc.). 
As an example, the 10\% error in the energy determination of cascades
which have the energy spectrum $\propto E^{-3}$ results in a systematic
overestimation of energies about 1\%. The corresponding shift of $X_{max}$
is less than 1 $gcm^{-2}$ and is negligible. The shift in $N_{max}$
is more dangerous since the difference between $N_{max}$ for different
nuclei is small. The whole 2-dimensional diagram shown in Figure 
\ref{fig:mta1}a will be displaced down by 60 - 70. 
This will reduce the abundance of light nuclei
and increase the abundance of heavy nuclei. The estimates made 'on the
back of envelope' show that the mean $\langle$ lnA $\rangle$ for the uniform 
primary mass composition can increase from 2.04 up to 2.20 - 2.25. That is why
the use of all the available information for the energy determination including
that from the surface detector is of crucial importance.\\
As for an estimation of uncertainty introduced from different hadronic
interaction model, if the real atmospheric cascades develop according to the 
QGSJET model, but analysed using another model, for example SIBYLL, 
then since the latter gives cascades slightly displaced towards deeper 
atmosphere, one can expect the shift of $\langle$ lnA $\rangle$
towards a heavier mass composition of the same magnitude as that caused by
an uncertainty of the energy determination.\\
Coming back to the methods for the determination of the primary mass
composition, certainly in order to use FCC the experimentalists should
have a set of mean cascade curves and their errors of the mean
for test cascades of different energies and zenith angles. 
Application of MTA and MMM is straightforward - the relevant simulations 
should be made with a maximum possible statistics.
The advantage of MTA is that it is very simple and easy to use. 
Also there is no problem to extend MTA for larger number of observables - 
one has just to inrease simulation statistics to create the relevant matrix.\\ 
For NNA the problem is more complex since
the capability of the neural network to identify particles is strongly 
related to the possibility to have a realistic T/V/T data sets. 
Therefore once these sets of data will be available, extensive testing will 
be performed aimed at selecting the most suitable neural model (in terms of
architecture, number of input neurons and hidden layers,
activation and error function etc.). One advantage of NNs is that
they may be optimized to work on well defined bins of the input
parameter space. The use of unsupervised NNs may also be of some
help in reducing the dimensionality of the parameter space.\\
Final goal of the efforts is to implement an optimal classifier
(Hyerarchical neural net or something else) capable to combine the 
classification of all methods tested so far.  

\section{Conclusions}

We proposed and tested four methods for the determination of the primary cosmic
ray mass composition in the EeV energy region on the basis of measurements of 
their longitudinal development: two of them (FCC and MTA)
are able to derive the mean mass composition and the other two 
(MMM and NNA) are based on the identification of the primary mass for each
individual cascade. However we stress that this classification depends 
essencially on the way to derive the probabilities
(except for FCC that can be really used only to derive mean mass composition).
In fact for MTA  we can assign a threshold to the
probabilities to associate each event with {\em just one} primary particle.
On the contrary, for MMM and NNA it can be possible to integrate over all 
showers to derive the mean mass composition. Among the proposed methods 
FCC ('Fit of the Cascade Curve'), MTA ('Multiparametric Topological
Analysis') and NNA ('Neural Net Analysis') with conjugate gradient 
optimization algorithm give the best accuracy.
All methods employ more information about the longitudinal development of 
atmospheric cascades than just the depth of the cascade maximum. 
They are independent and can be used complementary for the cross-check of 
final results.

\noindent {{\bf Acknowledgements}}

\noindent Authors thank Dr.M.Risse for the high statistics
simulation of cascades used in this work and remarks on the paper and also 
Prof. K.-H.Kampert, Prof. H.Rebel, Dr. L.Perrone
and Dr. V.P.Pavluchenko for remarks and useful discussions. One of the authors
(ADE) thanks INFN, Sezione di Napoli and the University of Catania for 
providing the financial support for this work and the hospitality.

\end{document}